# In Vivo Wideband MR Elastography for Assessing Age-Related Viscoelastic Changes of the Human Brain

Jakob Schattenfroh[1], Tom Meyer[1], Hossein S. Aghamiry[1], Noah Jaitner[1], Michael Fedders[1], Steffen Görner[1], Helge Herthum[2], Stefan Hetzer[2], Melanie Estrella[1], Guillaume Flé[3,4], Paul Steinmann[5], Jing Guo[1], Ingolf Sack[1*]

1. Department of Radiology, Charité - Universitätsmedizin Berlin, Berlin, Germany
2. Berlin Center for Advanced Neuroimaging, Charité - Universitätsmedizin Berlin, Berlin, Germany
3. Institute of Radiology, Universitätsklinikum Erlangen, Friedrich-Alexander-Universität Erlangen-Nürnberg, Erlangen, Germany
4. Institute of Neuroradiology, Universitätsklinikum Erlangen, Friedrich-Alexander-Universität Erlangen-Nürnberg, Erlangen, Germany
5. Institute of Applied Mechanics, Friedrich-Alexander Universität Erlangen-Nürnberg, Erlangen, Germany

# Abstract

Magnetic Resonance Elastography (MRE) noninvasively maps brain biomechanics and is highly sensitive to alterations associated with aging and neurodegenerative disease. Most implementations use a single frequency or a narrow frequency band, limiting the analysis of frequency-dependent viscoelastic parameters.

We developed a dual-actuator wideband MRE (5-50 Hz) protocol and acquired wavefields at 13 frequencies in 24 healthy adults (young: 23-39 years; older: 50-63 years). Shear wave speed (SWS) maps were generated as a proxy for stiffness, and SWS dispersion was modeled using Newtonian, Kelvin-Voigt, and power-law rheological models.

Whole-brain stiffness declined with age, with the strongest effect observed at low frequencies (5-16 Hz: -0.24%/year; p=0.019) compared with mid (20-35 Hz: -0.12%/year; p=0.030) and high frequencies (40-50 Hz: -0.10%/year; p=0.165). Compared to older brains, younger adults showed 14.3% higher baseline stiffness in the power-law model (p=0.001) and 8.5-9.0% higher viscosity according to the Newtonian and Kelvin-Voigt model (p<0.05). White and cortical gray matter exhibited similar age-related decreases, while deep gray matter showed an increase in the power-law exponent (+0.001/year; p=0.036), suggesting a transition toward more fluid-like properties associated with aging.

Wideband MRE revealed frequency-dependent and region-specific biomechanical alterations with aging, with the strongest effects observed at low frequencies. Extending brain MRE into the low frequency regime potentially enhances sensitivity to solid–fluid interactions. Therefore, low frequency MRE may serve as an early biomechanical marker of microstructural brain changes due to aging and neurodegeneration.

## Keywords

# 1.  Introduction

Magnetic Resonance Elastography (MRE) is a noninvasive imaging technique that quantifies the mechanical properties of soft biological tissues in vivo by analyzing their response to externally induced shear waves [1]. In the brain, MRE has shown promise for detecting mechanical changes associated with aging [2,3] and neurodegenerative disorders [4], such as Alzheimer's disease [5,6], Parkinson's disease [7,8], neuroinflammation [9,10], and tumors [11–14]. These mechanical biomarkers offer rich information for detecting, staging, and monitoring neurological disorders.

Most MRE methods use either a single vibration frequency [9,15,16] or a narrow frequency band (typically 30–60 Hz) [17,18], which limits the ability to assess viscoelastic dispersion, i.e., the variation of stiffness across vibration frequencies. Viscoelastic dispersion provides valuable insights into how different tissue compartments contribute to mechanical behavior and may be particularly sensitive at low excitation frequencies [19–21]. The brain has been shown to be much softer at 5 Hz than at 50 Hz, likely because mechanically active tissue structures respond differently to shear vibrations across the MRE frequency spectrum [22]. However, the lower frequency range, especially ultra-low frequencies below 10 Hz, remains largely unexplored, leaving in vivo MRE insensitive to slow, frequency-dependent processes.

At low and ultra-low actuation frequencies as well as under intrinsic pulsation-induced vibrations [23–26], poroelastic fluid-solid interactions, such as hydraulic conductivity, become increasingly relevant [24,27]. The poroelasticity framework helps explain why fluid turnover in brain tissue through glymphatic flow [28–31] or microvascular perfusion [32–34] affects brain stiffness, and why these effects are likely more pronounced at lower than at higher actuation frequencies. In vivo mapping of brain stiffness at ultra-low frequencies is, however, technically challenging due to the long shear wavelengths and small tissue strains, which reduce inversion sensitivity and make robust stiffness reconstruction more difficult [35,36]. As a result, this frequency range remains largely unexplored in human brain MRE, leaving a gap in our understanding of how frequency-dependent mechanical properties reflect the underlying physiology.

To overcome this limitation and probe stiffness dispersion across a wide frequency range in the human brain, we developed a dual-actuator MRE driver system. This system combines a lateral shear-wave excitation device for low frequencies (5-20 Hz) with a head-cradle actuator for higher frequencies (20-50 Hz). Data acquisition and post-processing of the 5-20 Hz band were optimized to avoid phase wrapping artifacts and prevent their propagation into the phase-gradient inversion [37,17]. Using voxel-wise rheological modeling, we assess global and regional frequency-dependent viscoelastic properties in healthy adults and evaluate their sensitivity to age-related changes.

Using this wideband MRE approach, we investigate age-related changes in viscoelastic dispersion of the healthy human brain. We hypothesize that the different frequency ranges accessed by wideband brain MRE provide distinct sensitivities to aging, suggesting possible clinical applications that leverage externally induced shear waves at ultra-low frequencies.

## 2. Methods

### 2.1. Study Design

This cross-sectional observational study included 24 healthy volunteers (6 females, 18 males; age (mean±SD): 40.4±13.8 years; range: 23-63 years) with no history of neurological disorders or head trauma. Participants were assigned to one of two age groups: younger adults (n = 14; 4 females, 10 males; mean age: 29.7±4.1 years; range: 23-39 years) and older adults (n = 10; 2 females, 8 males; mean age: 56.5±3.8 years; range: 50-63 years).

The study was approved by the institutional ethics review board. Written and oral informed consent was obtained from all participants prior to their inclusion.

### 2.2. Hardware Setup

A custom dual-actuator system (Fig. 1) was developed to enable wideband brain MRE across a frequency range of 5-50 Hz while fitting within a standard 32-channel head coil (Siemens, Erlangen, Germany). The compressed-air actuator system combined (Fig. 1A) a lateral driver composed of two air cushions (WANYIG QN-3 Mounting Cushion, Shenzhen, P.R. China) positioned symmetrically near the temples, and (Fig. 1B) a head cradle placed beneath the head and driven by two small plastic bottles [17]. The lateral driver was developed to enable low frequency vibrations down to the unprecedented range of 5 Hz. However, since this type of actuator has limitations in handling higher frequency vibration amplitudes (>20 Hz), it was combined with the established cradle driver which has proven effective in the frequency range of 20 to 50 Hz [17]. Both systems were powered by alternating air pulses to induce lateral head motion. The lateral driver (Fig. 1A) operated at low and ultra-low frequencies (5, 8, 10, 12.5, 16, 20 Hz), while the cradle driver (Fig. 1B) targeted higher frequencies (20, 25, 30, 35, 40, 45, 50 Hz). Frequency switching between acquisitions was performed automatically across the full range without repositioning the subject. Figure 1 illustrates the setup including the lateral placement of the air cushions and the position of the cradle driver beneath the head. Arrows indicate the alternating actuation pattern that ensures predominantly lateral head motion.

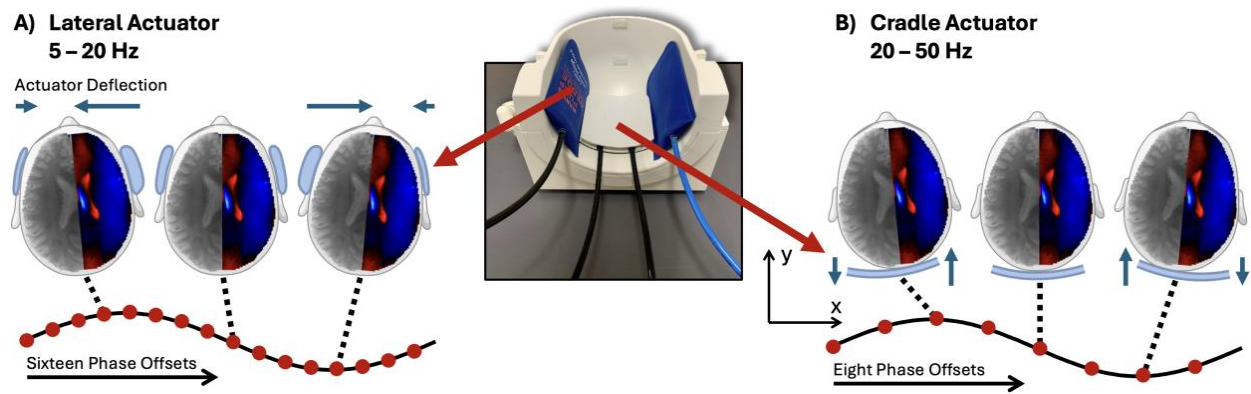

**Figure 1. Dual-Actuator Wideband MRE Setup.** Sketch and photograph of the dual-actuator system for wideband MRE combining: (A) a lateral driver based on two air cushions (blue), placed near the temples to induce ultra-low and low-frequency vibrations (5-20 Hz), and (B) a cradle driver positioned beneath the head to generate higher-frequency vibrations (20-50 Hz). Both actuators were connected to a pressurized air source via tubes (blue and black). To account for the longer vibration cycle duration at frequencies below 20 Hz, 16 dynamic wave phases were sampled per cycle in setup (A), compared to 8 phases for higher frequencies in (B). The subject and actuators remained in the same position across all frequencies.

## 2.3. Imaging Protocol

MRE was performed on a 3T clinical MRI scanner (Siemens Magnetom Lumina, Erlangen, Germany) using a single-shot spin-echo echo-planar imaging (SE-EPI) sequence with flow-compensated motion-encoding gradients (MEGs). The MEG amplitude was 34 mT/m with a slew rate of 125 mT/m/ms, durations of 19.8 ms for the 5-20 Hz range and 22.8 ms for the 20-50 Hz range. Imaging parameters were: field of view (FOV) 220×220×30 mm$^3$; matrix 136×136×15; spatial resolution 1.6×1.6×2 mm$^3$. Echo time (TE) was 56 ms (5-20 Hz) and 62 ms (20-50 Hz). Repetition time (TR) was 1400 ms (5-20 Hz) and 1140 ms (20-50 Hz). GRAPPA parallel imaging (acceleration factor 2) was used, and the receiver bandwidth was 1470 Hz/pixel. Vibrations were initiated 5 s before acquisition to ensure a steady state of harmonic oscillatory motion. Three-dimensional displacement fields were encoded consecutively along all three Cartesian axes of the imaging volume. To account for the extended duration of vibration cycles at lower frequencies (5-20 Hz), 16 equidistant dynamic wave phases were recorded over each vibration period. For higher frequencies (20-50 Hz), 8 phase offsets were used as in previous work [17]. The total MRE scan time across both frequency ranges was 10:46 min.

## 2.4. Data Processing

Slice-wise 2D motion correction was applied to the complex-valued MRE images [38] using SPM-realign [39]. EPI-distortion correction was performed using HySCO 2.0 [40] available in SPM12 [41]. Shear wave speed (SWS, m/s) as a surrogate for tissue stiffness was reconstructed from the phases of the complex MRE data using the publicly available *k*-MDEV inversion algorithm [37,42] (bioqic-apps.charite.de; brain pipeline [17,20]), with the default parameter setting for the higher frequency range (directional filtering of the wave field in eight propagation directions; Butterworth 3rd-order filter with passband from 15 m$^{-1}$ to 250 m$^{-1}$). For ultra-low and low frequencies (5-20 Hz), the inversion pipeline was adapted as follows (further details in the Appendix):

(i) Dense temporal sampling (16 wave phases) supported gradient-based phase unwrapping along the time dimension. Applying this method voxel-wise greatly reduced the number and spatial extent of phase wraps without the spatial bias often introduced by smoothness-enforcing unwrapping methods, particularly at low frequencies [43]. Remaining phase discontinuities were then removed by 2D Laplacian-based phase unwrapping [35,44].
(ii) Directional filtering of the wave field was applied along four in-plane directions using a cos$^2$ filter kernel [45,37], resulting in monodirectional shear waves propagating along the ±x and ±y axes within the imaging plane.
(iii) A linearly adaptive high-pass filter threshold [20] was used to separate low wave-number shear waves from compression waves.
(iv) Phase-gradients for *k*-MDEV inversion were derived only along the directions transverse to the polarization of the filtered waves to avoid introducing noise from non-shear wave components.

## 2.5. Viscoelastic Dispersion Analysis

The viscoelastic frequency dispersion was analyzed by fitting three models to frequency-resolved SWS data either (i) averaged over the brain tissue covered by MRE or (ii) voxel-wise to generate parameter maps. Viscoelastic dispersion was analyzed by fitting the Newtonian model, the Kelvin-Voigt model, and the spring-pot model (Table 1). The complex shear modulus $G^*$ was converted into SWS ($c$) based on the real and imaginary parts of $G^* = G' + iG''$, i.e., the storage modulus ($G'$) and loss modulus ($G''$) using [46]:

$$c(\omega) = \sqrt{\frac{2[G'^2 + G''^2]}{\rho\left[G' + \sqrt{G'^2 + G''^2}\right]}}. \tag{1}$$

**Table 1: Dispersion models.** Here $\omega$ is the angular vibration frequency, $\mu$ is stiffness, $\alpha$ is the spring-pot power-law exponent, and $\rho$ is the tissue density (assumed constant $1000\,\text{kg/m}^3$). For the spring-pot model, the reduced two-parameter fit $c(\omega) = c_0 \omega^{\alpha/2}$ is used (Appendix B).

| Model | $G^*$ | $c(\omega)$ | Fit variables (dimension) |
|---|---|---|---|

| Newtonian | $i\omega\eta$ | $\sqrt{\dfrac{2\eta\omega}{\rho}}$ | $\eta$ (Pa·s) |
|---|---|---|---|
| **Kelvin-Voigt** | $\mu + i\omega\eta$ | $\sqrt{\dfrac{2[\mu^2 + [\omega\eta]^2]}{\rho[\mu + \sqrt{\mu^2 + [\omega\eta]^2}]}}$ | $\mu$ (Pa), $\eta$ (Pa·s) |
| **Spring-Pot** | $\mu^{1-\alpha}\eta^{\alpha}[i\omega]^{\alpha}$ | $c_0\omega^{\alpha/2}$ | $c_0$ (m/s), $\alpha \in [0,1]$ |

Fitting was performed in MATLAB (R2024a; The MathWorks, Natick, MA, USA) using the $fminsearch$ function to minimize mean-square error. All SWS and viscoelastic parameter maps were normalized to the Montreal Neurological Institute (MNI) ICBM152 atlas space [47] using SPM12-normalize [41] based on the MRE magnitude images averaged over time steps, motion-encoding directions, and frequencies.

## 2.6. Statistical Analysis

Statistical analyses were performed in MATLAB. Group differences in SWS and dispersion-model parameters were assessed with the Wilcoxon rank-sum test for independent samples. Linear regression modeled the relationship between age and (i) SWS and (ii) dispersion-model parameters. Linear trends were evaluated using the coefficient of determination $R^2$, Pearson correlation coefficient PCC, and p-values. Statistical significance was set at p<0.05. Because voxel-wise fitting of rheological models is sensitive to noise and to single-voxel errors in frequency-resolved SWS maps, a cut-off threshold of $R^2 \geq 0.75$ for the fitted viscoelastic model was applied. Only voxels with $R^2$ at or above this threshold were used for further voxel-wise analysis. White matter (WM) and gray matter (GM) segmentation masks were defined through the maximum probability based on the tissue probability maps [48]. Deep gray matter (DGM) was defined based on the following regions of the Neuromorphometrics atlas (www.neuromorphometrics.com) included in SPM12: Accumbens, Amygdala, Caudate, Globus Pallidus, Putamen, Substantia Nigra, Thalamus.

# 3. Results
## 3.1. Shear Wave Propagation and Frequency-Resolved Elastograms

Figure 2 shows frequency-resolved shear wave fields after unwrapping and temporal Fourier transformation (real-valued $x$-components; subset of the full field) as well as SWS maps in a central transverse slice of a volunteer across the entire frequency range. The wavefields show distinct shear wave propagation patterns at each frequency, while lower frequencies exhibit longer wavelengths, higher frequencies demonstrate shorter wavelengths and increased wave damping toward central areas. SWS maps reflect viscoelastic dispersion of brain tissue by increasing values towards higher frequencies. The apparent SWS values observed in CSF-filled ventricles (red arrow) at high frequencies likely arise from turbulence and oscillatory fluid motion rather than true shear wave propagation, reflecting frequency-dependent fluid dynamics rather than intrinsic tissue stiffness.

Shear waves at 20 Hz were induced by both actuator systems. The resulting shear wave speeds showed no significant difference (lateral actuator: 1.209±0.066 m/s; cradle actuator: 1.219±0.058 m/s; Wilcoxon signed-rank test p=0.397), indicating that both driver types produce comparable shear wave speed maps at this overlapping frequency and can therefore be used interchangeably without introducing systematic bias in stiffness estimation.

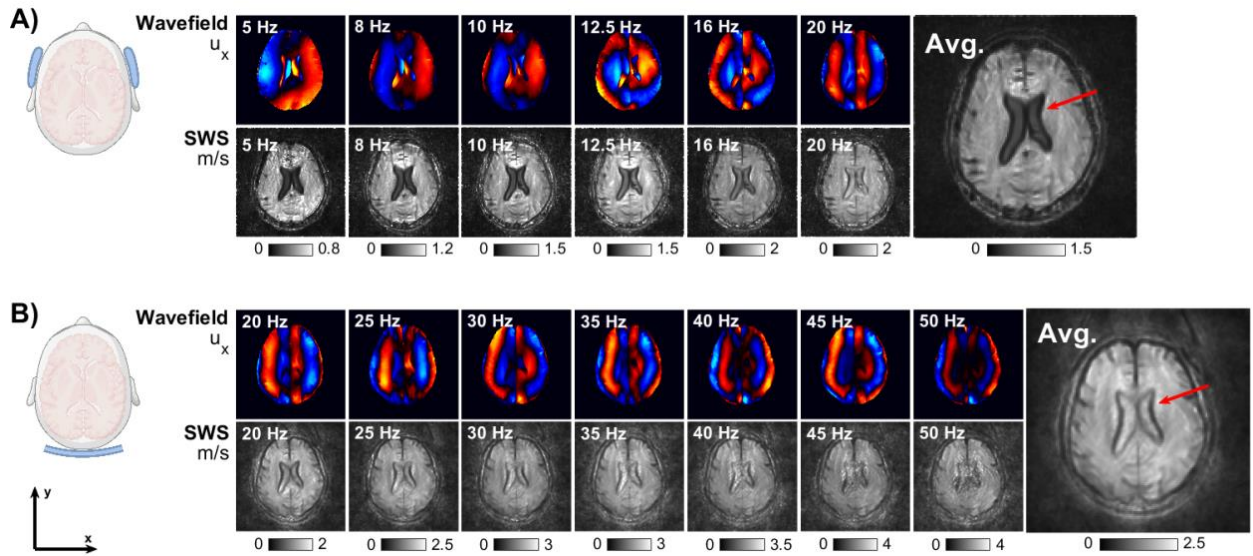

**Figure 2: Frequency-Resolved Wavefields and Shear Wave Speed Maps** (A) Lateral actuator at low frequencies (5, 8, 10, 12.5, 16, and 20 Hz). (B) Cradle actuator at high frequencies (20, 25, 30, 35, 40, 45, and 50 Hz). Top rows: axial wavefield slice with wave propagation along the x-direction and deflection along the y-direction. Bottom rows: corresponding frequency-resolved SWS maps. Averaged SWS maps for both frequency brackets are shown on the right.

### 3.2. Age-Related Differences in Brain Stiffness

Group-wise analysis of whole-brain SWS across two age cohorts (young: < 40 years; old: ≥ 40 years) revealed significant age-related differences in brain stiffness across frequency brackets (lower frequencies: 5-16 Hz; mid frequency range: 20-35 Hz; higher frequencies: 40-50 Hz; all: 5-50 Hz). Young subjects exhibited SWS values ranging from 0.74±0.06 m/s at lower frequencies to 2.03±0.09 m/s at high frequencies, whereas the older cohort showed values from 0.68±0.04 m/s to 1.94±0.08 m/s, respectively (p<0.05 for all frequency ranges). Linear regression analysis revealed an SWS decrease of 0.19% per year (p=0.019). Frequency-resolved analysis showed that lower and mid-range frequency SWS decreased with 0.242% (p=0.019) and 0.120% (p=0.030) per year, respectively. Higher frequency SWS was not significantly altered (0.10%/year decline; p=0.165), suggesting that MRE-measured brain softening is dominated by lower frequencies. Figure 3 shows the distribution of mean SWS values over age and across frequencies with linear model fits and 95% confidence intervals. Table 2 summarizes mean SWS values and linear regression parameters.

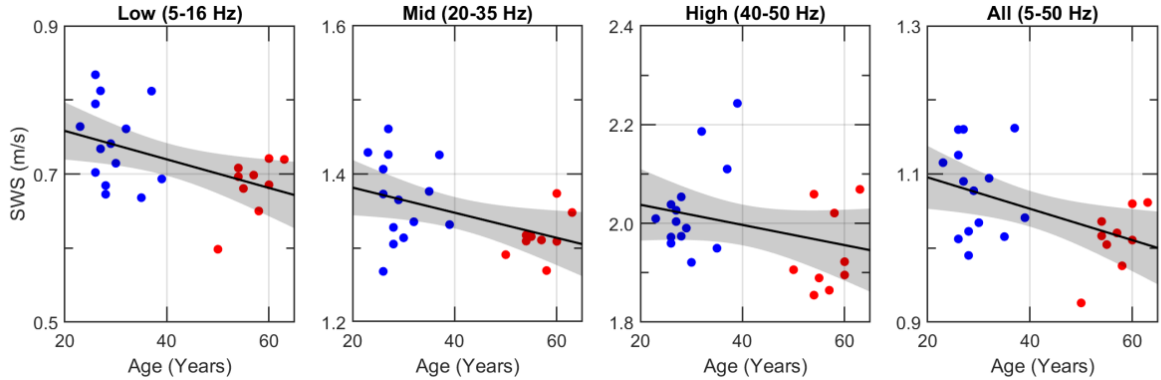

**Figure 3: Age-Related Changes in Averaged Shear Wave Speed (SWS).** Distribution of mean SWS values across frequency ranges for younger (23-39 years, blue dots) and older (50-63 years, red dots) subjects, with fitted linear regression models and 95% confidence intervals, showing a significant decrease in SWS with age. Model parameters are given in Table 1.

**Table 2: SWS Across Frequency Ranges and Age Groups.** SWS (mean±SD) for younger (23-39 years) and older (50-63 years) groups across different frequency ranges (Low: 5-16 Hz; Mid: 20-35 Hz; High: 40-50 Hz; All: 5-50 Hz). Linear regression models show the relationship between SWS and age, with equivalent annual percentage reductions based on model intercept and slope, $R^2$ values, PCC values, and p-values indicating statistically significant differences between the groups.

| Freq. Range | Group Mean SWS (m/s) | SWS Regression Model | $R^2$ | PCC | p |
|---|---|---|---|---|---|
| **Low** (5-16 Hz) | Young: 0.74±0.06<br>Older: 0.68±0.04 | 0.797-0.002×age<br>equivalent: -0.242% / year | 0.235 | -0.485 | 0.019 |
| **Mid** (20-35 Hz) | Young: 1.37±0.06<br>Older: 1.32±0.03 | 1.416-0.002×age<br>equivalent: -0.120% / year | 0.204 | -0.452 | 0.030 |
| **High** (40-50 Hz) | Young: 2.03±0.09<br>Older: 1.94±0.08 | 2.079-0.002×age<br>equivalent: -0.098% / year | 0.090 | -0.299 | 0.165 |
| **All** (5-50 Hz) | Young: 1.08±0.06<br>Older: 1.01±0.04 | 1.138-0.002×age<br>equivalent: -0.186% / year | 0.236 | -0.486 | 0.019 |

### 3.3. Group-averaged viscoelastic dispersion changes

Analysis of viscoelastic dispersion revealed age-related changes in rheological model parameters. Figure 4A and 4B show SWS dispersion as boxplots with fitted rheological model curves and standard deviations of the fits. SWS increased with frequency in both age groups, reflecting marked viscoelastic dispersion. However, the SWS increase was less pronounced in the older than the younger cohort. For the Newtonian model, η = 7.18±0.42 Pa·s in the young cohort compared to η = 6.57±0.36 Pa·s in the older cohort (-8.50%; p=0.004). For the Kelvin-Voigt model, η = 8.06±0.53 Pa·s and μ = 187.84±57.93 Pa in young subjects versus η = 7.38±0.48 Pa·s (-8.44%; p=0.009) and μ = 170.91±35.75 Pa (-9.01%; p=0.219) in older subjects. These results indicate the dominating role of viscosity over shear modulus in age-related biomechanical property changes toward lower viscosity in the elderly. The Kelvin-Voigt ($R^2$=0.725, p<0.001) and power-law ($R^2$=0.537, p=0.004) models best fit the SWS dispersion data. However, the power-law exponent α did not change significantly with age (young: 0.95±0.04; older: 0.97±0.01; +2.06%; p=0.123), while SWS decreased from $c_0$ = 0.14±0.01 m/s to $c_0$ = 0.12±0.0035 m/s (-14.29%; p=0.001) in the younger and older groups, respectively. Table 3 details these group comparisons.

Figure 4C shows the relative differences in acquired SWS and fitted rheological models across frequency. The largest SWS difference was 10.9±2.9% at 7 Hz and the lowest was 3.0±1.1% at 25 Hz. Difference-SWS according to the Newtonian model was constant across the full frequency range with 4.5±2.3%. Similarly, the Kelvin-Voigt model predicted only a minor dependence at very low frequencies and converged to a constant 4.4±2.6%. In contrast, the power-law model best predicted the increased age-difference of SWS observed at lower frequencies and showed the least deviation from difference SWS among all models (Newtonian model: 1.87%; Kelvin-Voigt model: 1.77%; power-law model: 1.58%).

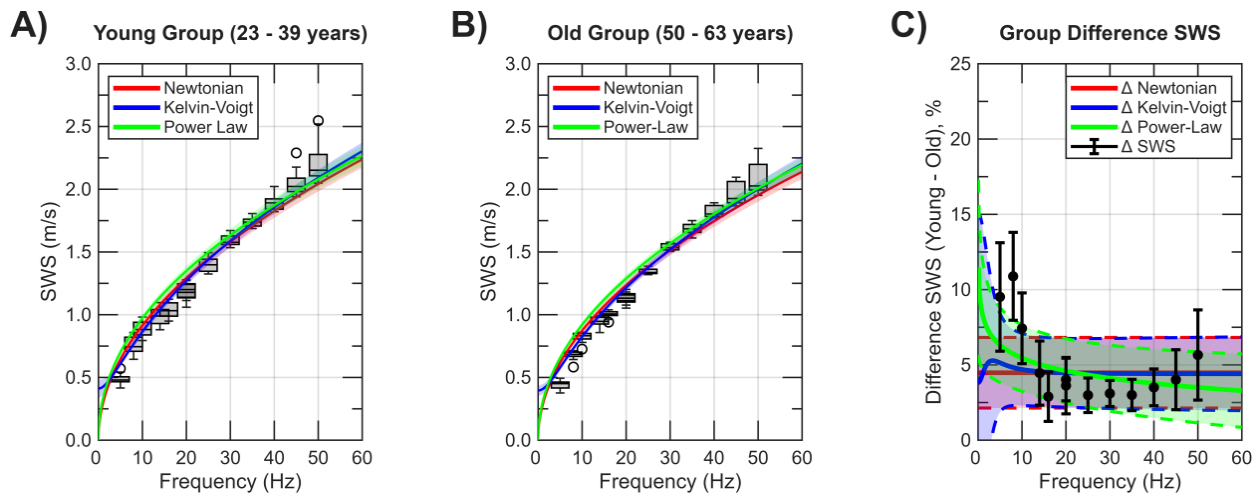

**Figure 4: Shear Wave Speed (SWS) Dispersion and Viscoelastic Modelling by Age Group.** Boxplots of the SWS dispersion curves for (A) younger and (B) older cohorts, fitted with group-averaged Newtonian, Kelvin-Voigt, and power-law rheological models. The curves highlight age-related differences in viscoelastic properties across the 5-50 Hz frequency range, with colored bands representing the standard deviation of fitted models across subjects. (C) Relative percent differences between SWS of younger and older groups (black) and fitted rheological models (colored).

**Table 3: Rheological Model Parameters for Young and Older Cohorts.** Parameters of Newtonian, Kelvin-Voigt, and power-law rheological models fitted to SWS data, comparing viscosity (η, Pa·s), shear modulus (μ, Pa), baseline SWS ($c_0$, m/s), and power-law exponent (α) between cohorts. Percent differences and p-values indicate statistical significance of group differences.

|  | **Newtonian** $G^* = i\omega\eta$ | **Kelvin-Voigt** $G^* = \mu + i\omega\eta$ | | **Power-Law** $c = c_0\omega^{\alpha/2}$ | |
|---|---|---|---|---|---|
|  | η in Pa·s | η in Pa·s | μ in Pa | $c_0$ in m/s | α |
| **Young** | 7.18±0.42 | 8.06±0.53 | 187.84±57.93 | 0.14±0.01 | 0.95±0.04 |
| **Old** | 6.57±0.36 | 7.38±0.48 | 170.91±35.75 | 0.12±0.0035 | 0.97±0.01 |
| **% Diff** | -9.0% | -8.8% | -9.4% | -10.8% | 2.5% |
| **p** | 0.004 | 0.009 | 0.219 | 0.001 | 0.123 |

### 3.4. Individual viscoelastic parameter changes with age

Linear regression analysis (Figure 5; Table 4) of individual viscoelastic model parameters showed a viscosity decrease with age according to both the Newtonian model (slope: -0.018 Pa·s/year; p=0.013) and the Kelvin-Voigt model (slope: -0.018 Pa·s/year; p=0.043), while shear modulus showed no significant change (p=0.959). This again highlights the sensitivity of viscous damping in age-related brain mechanical property changes, indicating a shift toward less viscous behavior in the elderly. The power-law model indicated a significant decrease in baseline SWS ($c_0$; slope: -0.484×10$^{-3}$ m/s/year; p=0.007) and a small but significant increase in the power-law exponent α (slope: 0.873×10$^{-3}$ 1/year; p=0.048). Together, these findings suggest that the ratio of absolute viscosity to shear modulus across the whole brain remains largely constant with age, implying a preserved relative viscosity or consistent fluid-like mechanical behavior, despite changes in absolute parameters.

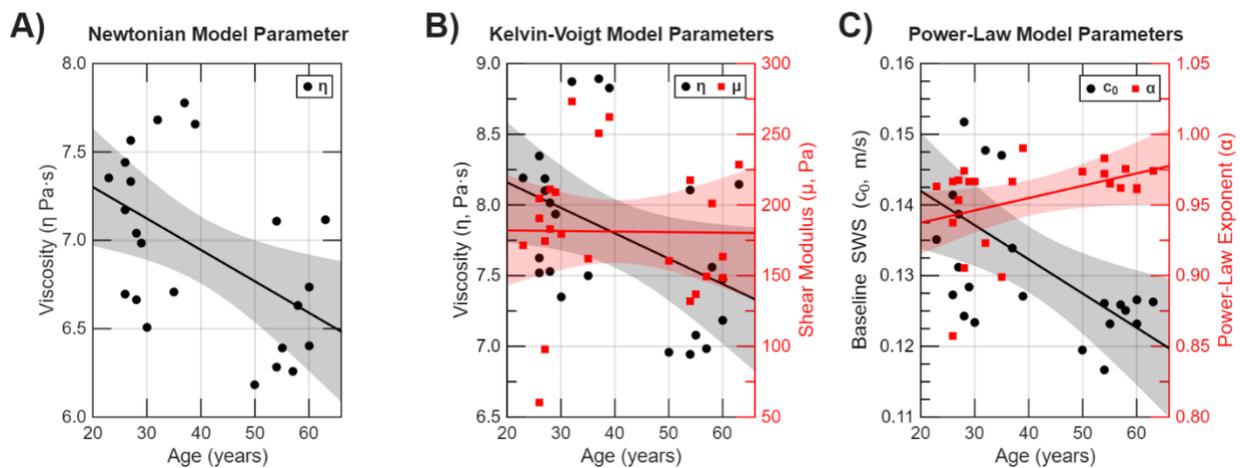

**Figure 5: Age-Resolved Viscoelastic Model Parameters.** Linear trends and confidence intervals of viscoelastic parameters (viscosity η, shear modulus μ, baseline SWS $c_0$, and power-law exponent α) as a function of age, derived from (A) Newtoniann, (B) Kelvin-Voigt, and (C) power-law model fits of frequency-resolved SWS data.

**Table 4: Age-Resolved Trends in Rheological Model Parameters.** Linear regression analysis of viscoelastic parameters (viscosity η, shear modulus μ, baseline SWS $c_0$, and power-law exponent α) as a function of age, derived from Newtonian, Kelvin-Voigt, and power-law model fits. Slope, offset, $R^2$, PCC, and p-values indicate the strength and significance of age-related trends.

|  | Newtonian $G^* = i\omega\eta$ | Kelvin-Voigt $G^* = \mu + i\omega\eta$ | | Power-Law $c = c_0\omega^{\alpha/2}$ | |
|---|---|---|---|---|---|
|  | $\eta$ in Pa·s | $\eta$ in Pa·s | $\mu$ in Pa | $c_0$ in m/s | $\alpha$ |
| **Slope** | -0.018 | -0.018 | -0.040 | -0.484×10$^{-3}$ | 0.873×10$^{-3}$ |
| **Offset** | 7.657 | 8.519 | 182.804 | 0.152 | 0.920 |
| **R$^2$** | 0.261 | 0.181 | 0.000 | 0.301 | 0.161 |
| **PCC** | 0.511 | 0.425 | 0.011 | 0.549 | 0.401 |
| **p** | 0.013 | 0.043 | 0.959 | 0.007 | 0.048 |

## 3.5. Spatially Resolved Viscoelastic Property Changes with Age

Voxel-wise fitting of rheological models provided spatially resolved viscoelastic parameter maps. Figure 6A shows median parameter maps for all 24 subjects in a central slice of the MNI atlas, revealing distinct spatial patterns for all four parameters. An $R^2$ value ≥ 0.75 was observed in 89.2 ± 5.4% of voxels for the Newtonian model, 91.7 ± 4.2% Kelvin-Voigt model, and 88.0 ± 6.4% for the Power-Law model.

Group-wise comparisons between younger and older participants showed significant differences in several parameters. For the Newtonian model global brain tissue (GBT) viscosity was 7.43±0.35 Pa·s in young subjects and 6.85±0.37 Pa·s in older subjects (p=0.003), white matter (WM) was 8.13±0.28 vs. 7.69±0.25 Pa·s (p=0.003), cortical gray matter (GM) was 6.75±0.49 vs. 6.05±0.49 Pa·s (p=0.005), while deep gray matter (DGM) showed no significant difference (7.21±1.39 vs. 7.30±1.27 Pa·s; p=0.925). The Kelvin-Voigt model yielded similar results with GBT at 7.99±0.36 vs. 7.40±0.45 Pa·s (p=0.007), WM at 8.74±0.30 vs. 8.30±0.30 Pa·s (p=0.007), GM at 7.30±0.53 vs. 6.58±0.62 Pa·s (p=0.011), and DGM again showed no difference (7.38±1.43 vs. 7.62±1.17 Pa·s; p=0.682).

For the power-law model, baseline SWS $c_0$ was lower in older subjects in GBT (0.14±0.01 vs. 0.13±0.00 m/s; p=0.018) and GM (0.13±0.01 vs. 0.13±0.00 m/s; p=0.003), with no significant difference in WM (0.15±0.01 vs. 0.14±0.01 m/s; p=0.055) and DGM (0.15±0.02 vs. 0.14±0.02 m/s; p=0.270). The power-law exponent α did not differ significantly between age groups in GBT, WM, or GM, but showed a trend toward higher values in DGM (0.91±0.04 vs. 0.93±0.02; p=0.065).

Figure 6C illustrates these age-related changes in rheological parameters based on linear regression analysis. Slopes for the Newtonian model were -0.0172 Pa·s/year (p=0.008) in GBT, -0.0124 Pa·s/year (p=0.012) in WM, -0.0214 Pa·s/year (p=0.012) in GM, and 0.0015 Pa·s/year (p=0.941) in DGM. The Kelvin-Voigt model showed slopes of -0.0172 Pa·s/year (p=0.014) in GBT, -0.0118 Pa·s/year (p=0.027) in WM, -0.0218 Pa·s/year (p=0.022) in GM, and 0.0064 Pa·s/year (p=0.753) in DGM.

For the power-law model, $c_0$ declined with age in GBT (-0.0003 m/s/year, p=0.009), WM (-0.0003 m/s/year, p=0.021), and GM (-0.0003 m/s/year, p=0.005), while DGM showed no significant decline (-0.0005 m/s/year, p=0.142). The exponent α exhibited no significant age dependence in GBT, WM, or GM, but increased significantly with age in DGM (0.0010/year, p=0.036).

These findings agree with the whole-brain analysis in Figures 4 and 5, confirming the robustness of regional viscoelastic differences and their age-related trends. Both viscosity and $c_0$ decreased with age across most brain regions, whereas α remained constant except for a small but significant increase in DGM.

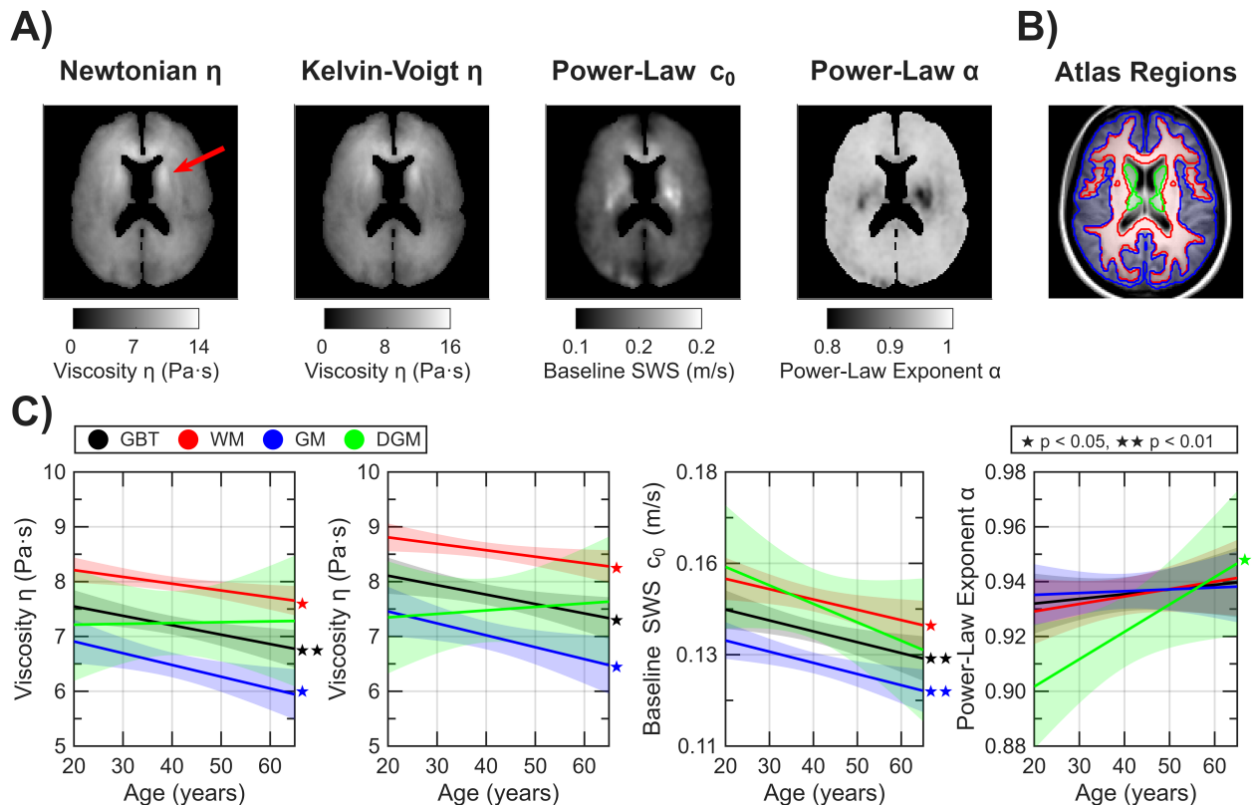

**Figure 6: Spatially Resolved Viscoelastic Parameter Maps and Age-related Changes.** (A) Spatially resolved maps of viscoelastic parameters derived from frequency-resolved SWS data, showing regional variations in biomechanical properties across the brain. The red arrow highlights the deep gray matter region, showing properties distinct from surrounding tissue. (B) Global brain tissue (GBT), white matter (WM), gray matter (GM), and deep gray

matter (DGM) segmentation masks in atlas space. (C) Fitted linear model with 95% confidence interval showing change in viscoelastic parameters over age for different atlas regions.

## 4. Discussion

Using a new dual-actuator system for in vivo brain MRE, we explored an unprecedented frequency range spanning more than three octaves (5-50 Hz). This work extends brain MRE into the ultra-low frequency regime (5-10 Hz) with spatially resolved SWS maps, bridging the gap between intrinsic pulsation (~1 Hz) studies [23–25,32,49], external actuation below 10 Hz without spatially resolved parameter mapping [20], and wideband MRE implementations in a more common frequency range of >10 Hz [35]. The previously unexplored band of ultra-low external vibration frequencies is potentially sensitive to fluid-solid interactions that dominate brain tissue mechanics at slow dynamics. In addition to blood-filled vessels and capillaries, brain tissue is saturated with interstitial fluid, CSF, and glymphatic channels - each contributing to viscous dissipation, particularly under low-frequency deformation [22,24,50]. Viscous loss manifests as attenuation of shear wave amplitudes (absolute viscosity) and frequency-dependent SWS dispersion (relative viscosity). These effects reflect microstructural solid-fluid interactions and may be sensitive to physiological processes such as glymphatic activity [28–30], age-related parenchymal atrophy [2,3,51,52], and progressive microvascular degeneration [18].

The lateral driver was optimized for low-frequency vibrations and the cradle driver provided stable excitation at higher frequencies. Their overlap at 20 Hz showed no significant difference in shear wave speed. This agreement confirms that both systems can be seamlessly integrated without bias, ensuring robust and reproducible stiffness measurements across the full frequency spectrum.

We observed a significant age-dependent decline in brain stiffness of approximately 0.24% per year in the low-frequency range (5-16 Hz), with a smaller and non-significant effect of -0.10% per year at higher frequencies (40-50 Hz). Across the full 5-50 Hz spectrum and consistent with previous studies [52], stiffness declined by 0.19% per year. Rheological analysis corroborated prior reports of (i) superviscous brain properties [20], (ii) reductions in shear modulus and absolute viscosity with age [53], and (iii) widely constant or a slight regional increase in relative viscosity (power-law exponent α, damping ratio, tissue fluidity) with age [52,54]. In our data, both the Newtonian and Kelvin-Voigt models revealed significant viscosity reductions, while shear modulus changes were not significant. Because power-law SWS is sensitive to viscosity (α), the lumped parameter $c_0$ decreased with age, likely driven by viscosity rather than stiffness. Spatially resolved rheological maps confirmed higher absolute viscosity in white matter and deep gray matter compared to cortical gray matter, consistent with prior single-frequency MRE studies [9,55]. Regionally, viscosity decreased with age in white matter, cortical gray matter, and global brain tissue, whereas deep gray matter viscosity remained relatively stable. Constant viscosity alongside decreasing stiffness with age in deep gray matter might explain our observation of increasing relative viscosity (power-law exponent $\alpha$) in that region.

The observed differences in absolute and relative viscosity provide complementary insights into brain tissue mechanics. Absolute viscosity, as captured by η in the Newtonian and Kelvin-Voigt models, reflects the magnitude of viscous damping and is strongly influenced by microstructural components such as myelin, extracellular matrix, and vasculature [22]. Its significant reduction with age suggests a loss of viscous dissipation capacity, potentially linked to demyelination, extracellular matrix remodeling, or reduced microvascular integrity. Relative viscosity, represented by the power-law exponent α, describes the frequency dependence of SWS and is sensitive to the balance between viscous and elastic contributions across scales. While α showed only subtle age-related trends, it may capture changes in viscoelastic dispersion that are not evident from absolute viscosity alone. Combined, these parameters offer a comprehensive biomechanical characterization: absolute viscosity highlights changes in energy dissipation, whereas relative viscosity reveals changes in the structural hierarchy of viscoelastic elements, leading to different scaling properties and solid-fluid phase transitions. Our study demonstrated that both properties change with age, with the strongest effects at lower frequencies. Therefore, low frequency MRE may offer enhanced sensitivity to (patho-) physiological alterations associated with solid-fluid interactions and attenuation, enabling early detection of microstructural deterioration in conditions such as small-vessel disease or neurodegeneration. The clinical feasibility of adding low-frequency MRE is supported by an acquisition time of 7:17 min for six frequencies ≤ 20 Hz.

This study is limited by the relatively small sample size. The lack of participants continuously covering the age spectrum supported our grouping approach into groups of young and elderly, but it assumes linear regression between them. Potential confounders such as cardiovascular health and hydration status were not controlled. Apparent SWS in CSF regions at low frequencies, reflecting fluid dynamics, underscores the need for careful segmentation. Furthermore, rheological model fits rely on assumptions of isotropy and homogeneous density, which may not fully capture poroelastic effects in aging brain tissue, while voxel-wise estimates are susceptible to partial volume effects and noise in small structures.

In summary, this study demonstrates the feasibility and relevance of wideband MRE for characterizing age-related changes in viscoelasticity. By extending the frequency range into the ultra-low regime below 10 Hz and combining MRE with voxel-wise rheological modeling, we show the dominant roles of absolute and relative viscosity, solid-fluid tissue interactions, and low-dynamic mechanical processes in the biomechanical characterization of physiological alterations of the in vivo human brain. Low-frequency MRE may provide a sensitive biomarker for aging and neurodegeneration in the clinical context.

# Funding

This work was supported by the German Research Foundation [grant numbers 460333672 CRC1540 EBM, RTG2260 BIOQIC, CRC1340 Matrix-in-vision, FOR5628, 540759292 Sa901/33-1 M5].

# Appendix A: Processing Pipeline

This appendix details the processing adaptations used to improve phase unwrapping and inversion robustness for MRE at ultra-low frequencies. The brain-specific *k*-MDEV inversion pipeline is outlined in [17,20], to which the following four adaptations were implemented:

**(i) Temporal Gradient Phase Unwrapping (TGPU). TGPU** resolves spatiotemporal phase discontinuities while preserving spatial wave integrity. At low frequencies, spatial unwrapping methods (e.g., Laplacian-based) struggle with long wavelengths, which can lead to wave-bending artifacts. Operating voxel-wise in the temporal domain, TPGU avoids these issues. In MRE, the phase signal at each voxel is modulated by harmonic tissue vibrations and is inherently sinusoidal but observed with wraps due to the intrinsic limitation of MRI phase measurements within $(-\pi, \pi]$. The true phase signal over the vibration at each voxel can be modeled as: $\varphi(t) = A \sin(2\pi f t + \varphi_0) + \varphi_{BG}$, where $A$ is the amplitude (proportional to maximum induced displacement), $f$ is the drive frequency of the applied vibration, $\varphi_0$ is the initial offset, $\varphi_{BG}$ the constant background phase offset, and $t$ is time. However, the measured phase, $\varphi_w(t)$, is wrapped: $\varphi_w(t) = \varphi(t) + 2\pi k$, where $k \in \mathbb{Z}$ such that $\varphi \in (-\pi, \pi]$. TGPU exploits harmonicity by computing the temporal phase gradient of the wrapped phase:

$$\Delta\varphi_w(t_i) = \arg\{\exp(i\varphi_w(t_{i+1})) \cdot \exp(-1i\varphi_w(t_i))\}, \qquad \text{Eq. (A.1)}$$

where $t_i$ are the discrete time points sampled over one vibration period, typically with $N$ equidistant phase offsets, such that $\Delta t = T/N = 1/(fN)$. Due to the periodic nature of the phase offsets one can assume $\varphi_w(t_{N+1}) = \varphi_w(t_1)$. For the true unwrapped phase, the difference approximates the derivative:

$$\Delta\varphi(t_i) \approx \left(\frac{d}{dt}\varphi\right)(t_i) \cdot \Delta t, \qquad \text{Eq. (A.1)}$$

where $(d\varphi/dt)(t) = A\omega \cos(\omega t + \varphi_0)$ being itself a sinusoidal function, shifted in phase by $\pi/2$ and scaled by $A\omega$, and importantly maintaining the harmonic nature of the signal. If the temporal sampling is sufficient (i.e., $N$ is large), the phase differences $\Delta\varphi_w(t_i)$ are small. The maximum phase differences can thus be determined by:

$$|\Delta\varphi|_{\max} \approx |A\omega| \cdot \Delta t = A \cdot \frac{2\pi}{N}. \qquad \text{Eq. (A.3)}$$

For sixteen timesteps (i.e., $N = 16$), $|\Delta\varphi|_{\max} = A \cdot \pi/8$. If $A < 8$, then $|\Delta\varphi|_{\max} < \pi$, ensuring that the phase differences lie within $(-\pi, \pi]$ which allows the assumption of $\Delta\varphi(t_i) = \Delta\varphi_w(t_i)$. Therefore, we can then reconstruct the unwrapped phase through $\varphi_{uw}(t_i) = \sum_{k=1}^{i-1} \Delta\varphi_w(t_i)$. If TGPU fails to fully resolve phase wraps, Laplacian-based unwrapping [35,44] is used as a secondary method. Although Laplacian-based unwrapping may

introduce wave-bending artifacts due to reliance on spatial gradients, it serves as a robust fallback to ensure complete phase unwrapping when temporal methods are insufficient.

**(ii) Harmonic Wave Field Extraction:** The unwrapped phase $\varphi_{uw}(t_i)$ is Fourier-transformed in the temporal domain to extract the harmonic wave field at the drive frequency $f$. The wave field is decomposed into four planar shear wave fields $\boldsymbol{u}_{d,f}(x,y,\theta_n)$ using a spatial directional filter with $\cos^2$ dependence [45], where $\theta_n, n \in \{\pm x, \pm y\}$ representing four propagation directions along the Cartesian axes of the imaging coordinate system, $d$ the motion-encoding directions, and $f$ are the driver frequencies.

**(iii) Frequency-Adaptive High-Pass Filtering.** A third-order high-pass Butterworth filter with a linearly increasing threshold (high-pass threshold $[\text{m}^{-1}] = 2 + 0.5f$, for $5 \leq f \leq 20$ Hz) isolates induced shear waves from noise and other low-frequency signals, such as compression waves.

**(iv) Phase-Gradients:** Due to alternating activation, (one cushion inflates while the other deflates), only lateral motion was induced, with the main shear wave propagation directions within the axial slice. Therefore, through-plane motion was neglected and only 2D axial derivatives were computed using finite-difference gradient kernels introduced by Anderssen and Hegland [56]. To calculate the real part of the complex wave number $k$, the 2D phase gradient was calculated along the respective propagation direction of the four planar shear wave fields (i.e., along $x$ for $u_{\pm x}$ and along $y$ for $u_{\pm y}$) for each motion encoding direction $d$ and driver frequency $f$:

$$k'_{d,n,f}(x,y,\theta_{n=\pm x}) = \frac{1}{2 \cdot resX} \arg\{\bar{\boldsymbol{u}}_{d,n,f}(x,y,\theta_{n=\pm x}) \cdot \boldsymbol{u}_{d,n,f}(x+1,y,\theta_{n=\pm x})\} \quad \text{Eq. (A.4a)}$$

$$k'_{d,n,f}(x,y,\theta_{n=\pm y}) = \frac{1}{2 \cdot resY} \arg\{\bar{\boldsymbol{u}}_{d,n,f}(x,y,\theta_{n=\pm y}) \cdot \boldsymbol{u}_{d,n,f}(x,y+1,\theta_{n=\pm y})\} \quad \text{Eq. (A.4b)}$$

Frequency-resolved SWS maps, $SWS_f$, were derived through weighted averaging of $k'_{d,n,f}$, while frequency-averaged $SWS$ values are computed using the harmonic mean of $SWS_f$:

$$SWS_f = \left[\frac{\sum_{d,n}|k'_{d,n,f}| \cdot |\boldsymbol{u}_{d,n,f}|^w}{\omega_j \cdot \sum_{d,n}|\boldsymbol{u}_{d,n,f}|^w}\right]^{-1}, \qquad SWS = \frac{N}{\sum_{f=1}^{N} 1/SWS_f} \qquad \text{Eq. (A.5)}$$

# Appendix B: Spring-Pot Model Dispersion

For the spring-pot viscoelastic model

$$G^*(\omega) = \mu^{1-\alpha}\eta^\alpha [i\omega]^\alpha = \kappa[i\omega]^\alpha = \kappa\omega^\alpha e^{i\frac{\pi\alpha}{2}}, \qquad \text{Eq. (B.1)}$$

where $\kappa = \mu^{1-\alpha}\eta^\alpha$ and $0 \leq \alpha \leq 1$. The real and imaginary parts are

$$G'(\omega) = \kappa\omega^\alpha \cos\left(\frac{\pi\alpha}{2}\right) \quad \text{and} \quad G''(\omega) = \kappa\omega^\alpha \sin\left(\frac{\pi\alpha}{2}\right). \qquad \text{Eq. (B.2)}$$

Substituting Eq. (B.2) in Eq. (1) (phase velocity defined via the real part of the complex wavenumber) gives

$$c(\omega) = \sqrt{\frac{2\kappa}{\rho\left[1 + \cos\left(\frac{\pi\alpha}{2}\right)\right]}} \, \omega^{\frac{\alpha}{2}}. \qquad \text{Eq. (B.3)}$$

Since $1 + \cos\left(\frac{\pi\alpha}{2}\right) = 2\cos^2\left(\frac{\pi\alpha}{4}\right)$, Eq. (B.3) can be written as

$$c(\omega) = \sqrt{\frac{\kappa}{\rho}} \frac{1}{\cos\left(\frac{\pi\alpha}{4}\right)} \, \omega^{\frac{\alpha}{2}} = c_0 \omega^{\frac{\alpha}{2}}, \qquad \text{Eq. (B.4)}$$

where $c_0 = \sqrt{\kappa/\rho}/\cos\left(\frac{\pi\alpha}{4}\right)$, with the lumped frequency-independent SWS coefficient $c_0$, which can take values between $\sqrt{\mu/\rho}$ for $\alpha = 0$ (elastic model) and $\sqrt{2\eta/\rho}$ for $\alpha = 1$ (Newtonian model). To avoid overfitting, we adopted the reduced two-parameter fit $c(\omega) = c_0 \omega^{\alpha/2}$, treating $c_0$ and $\alpha$ as independent parameters.